\documentstyle[12pt]{article}

\let\barr=\overline

\def\Ham{{\cal H}}
\def\E{{\cal E}}
\def\P{{\cal P}}
\def\half{\frac{1}{2}}
\def\xhat{{\widehat x}}
\def\phat{{\widehat p}}
\def\STvac#1{|#1\rangle_{\rm ST}}
\def\FSvac#1{|#1\rangle_{\rm FS}}

\def\talpha{\widetilde{\alpha}}

\def\tvac{\widetilde{|0\rangle}}

\def\mysection#1{\setcounter{equation}{0}
  \stepcounter{section}
  \section*{\large\bf\arabic{section}\ \  #1}
}

\begin{document}

\renewcommand{\thefootnote}{\fnsymbol{footnote}}
\renewcommand{\theequation}{\arabic{section}.\arabic{equation}}

\newcounter{saveeqn}
\newcommand{\alpheqn}%
{ \setcounter{saveeqn}{\value{equation}}%
  \stepcounter{saveeqn}\setcounter{equation}{0}%
  \renewcommand{\theequation}%
  {\mbox{\arabic{section}.\arabic{saveeqn}\alph{equation}}}}
\newcommand{\reseteqn}%
{ \setcounter{equation}{\value{saveeqn}}%
  \renewcommand{\theequation}{\arabic{section}.\arabic{equation}}}

%
%

\def\ben{\begin{equation}}
\def\een{\end{equation}}
\def\bea{\begin{eqnarray}}
\def\eea{\end{eqnarray}}
\def\nn{\nonumber}

\def\BU{\it
                {}\\
                {${}^1$}Department of Physics\\
                Boston University\\
                Boston MA\ \ 02215\\
                {\tt eben@budoe.bu.edu, hyuklee@buphy.bu.edu}}

\def\MIT{\it
                {}\\
                {${}^2$}Department of Physics\\
                Massachusetts Institute of Technology\\
                Cambridge MA\ \ 02139\\
                {\tt jackiw@marie.mit.edu}}


%
\hfuzz=5pt
\tolerance=10000
\begin{titlepage}
\begin{center}
{\Large Functional Schr\"odinger and BRST Quantization\\
of (1+1)--Dimensional Gravity}
\footnote
{This work was supported in part by funds provided by the U.S.
Department of Energy (D.O.E.) under contracts \#DE-FG02-91ER40676
and \#DE-FC02-94ER40818, and by Korea Science and Engineering Foundation
(KOSEF).}

\vspace{.6in}
{E.~Benedict,{${}^1$}
R.~Jackiw,{${}^2$} H.-J.~Lee{${}^1$}}
\vspace{.4in}
\BU
\vspace{.2in}
\MIT
\end{center}
\vspace{.8in}
\abstract{
We discuss the quantization of pure string--inspired dilaton--gravity in
$(1+1)$--dimensions, and of the same theory coupled to scalar matter. We
perform the quantization using the functional Schr\"odinger and
BRST formalisms. We find, both for pure gravity and the matter--coupled
theory, that the two quantization procedures give inequivalent
``physical'' results.
}
\vfill
\hbox to \hsize{BU-HEP 96-17\hfill}
\hbox to \hsize{MIT-CTP-2544\hfill July, 1996}
\end{titlepage}

\setlength{\baselineskip}{24pt}

\mysection{Introduction}

In four space--time dimensions, gravitational theories are
non--renormalizable, which makes their quantization problematic using
conventional field--theoretical techniques. In addition, there are
conceptual difficulties that are peculiar to diffeomorphism--invariant
quantum theories like
gravity. These include
the question of how to introduce time into
the theory, and the interpretational issues that arise from the
unfamiliar role of the Hamiltonian as a constraint. Recently, much work
has been done on gravitational theories in $(1+1)$--dimensions, in which the
computational difficulties that one faces in $(3+1)$--dimensions are
absent, because there are no propagating
gravitons in the lower dimensional theory,
and the question of renormalizability does not arise. The conceptual
issues remain, however, and the $(1+1)$--dimensional models are
useful for investigating them.

In two space--time dimensions the Einstein tensor vanishes identically, so
$(1+1)$--dimensional gravity models cannot be based on the
Einstein--Hilbert action. A variety of models have been proposed, the
most popular of which is the string--inspired CGHS model
\cite{CGHS,Verlinde}, which belongs to the class of scalar--tensor theories
introduced over a decade ago \cite{ScalTens}.
Our paper is concerned with a theory, related by
a conformal transformation
to the CGHS model, and described by
the action
\ben
S=\int dt\int d\sigma\> \sqrt{-g}(\eta R-2\lambda)\>.\label{StrInsp}
\een
This theory has been quantized, using both the metric--based action
(\ref{StrInsp}) \cite{Kuns}
and an equivalent gauge--theoretical action, invariant
under the extended Poincar\'e group \cite{Jack,Jack2}.
The quantization consists
of solving the Dirac constraints in a functional
Schr\"odinger formulation.

Recent work on the string--inspired
model \cite{CJZ} has shown that it is similar
to the theory of a bosonic string propagating on a $(1+1)$--dimensional
background space--time.
It is a familiar result from string theory that the bosonic
string cannot be straightforwardly
quantized on a two--dimensional target space, because
the Virasoro anomaly gives a center to
the algebra of constraints.  In
light of this, it is surprising that solutions to the quantum constraint
algebra derived from (\ref{StrInsp}) can be constructed
\cite{Kuns,Jack2}, since their
existence implies that the center vanishes.

The resolution of the apparent contradiction \cite{CJZ}
relies on the fact that the anomaly is present or absent
depending on the vacuum that is used to ``normal'' order the constraints:
the conventional vacuum of string theory is not appropriate for
the functional Schr\"odinger representation that was used in
\cite{Kuns,Jack2,CJZ}.

However, string--like theories in $(1+1)$--dimensions have been
constructed, using the BRST approach \cite{Polyakov,BMP,LZ,Bilal}.
They are made consistent through the addition of
background charges and ghost fields,
and the resulting spectrum contains more states than the ones
found in the gravitational theory: there is a continuous component,
consisting of two families of states labeled by the zero--mode momenta
of the fields;
in addition, there is an infinite tower of
``discrete states'' that
appear at special values of
the zero--mode momenta.

We shall show that the continuous component of the BRST spectrum
finds its analog in the functional Schr\"odinger
quantization when we solve the theory (\ref{StrInsp}) on the cylindrical
space--time ${\bf R}^1\times S^1$. The tower of
discrete states, on
the other hand, arises only when background charges are
present. Since there is
no center in the functional Schr\"odinger ordering, there is no need to
introduce background charges, and we find only two additional discrete
states with vanishing zero--mode momentum.

We shall present a similar comparison for the matter--coupled theory. We
use the results of \cite{CJZ} to
show that the action (\ref{StrInsp}), treated in the functional
Schr\"odinger formalism, and
minimally coupled to scalar matter,
is equivalent to an action studied in \cite{Kuch}. The theory cannot
be quantized without modification, due to an obstruction in the
matter field contribution to the constraint algebra. A
consistent quantum theory is constructed by adding terms to the
constraints that canonically
yield a center in the Poisson bracket algebra of
constraints, which cancels the center in the quantum theory
that arises from ordering the matter field contribution to the
constraint commutator algebra \cite{CJZ,Kuch}.
Furthermore, we consider the alternative BRST approach where the total
non--vanishing center arises from the gravitational and matter fields,
as well as
from background charges, and is canceled by the ghost contribution.

In the remainder of this Section, we review the transformations on
(\ref{StrInsp}) that bring out the analogy to $(1+1)$--dimensional string
theory. Also we discuss the origin of the anomalies in the algebra of
constraints and show that they are absent in the Schr\"odinger
representation, by recording explicitly the states on ${\bf R}^2$
that solve the constraints.

To put (\ref{StrInsp}) in
canonical form,
we parameterize the metric tensor $g_{\mu\nu}$ as in \cite{Kuns},
\ben
g_{\mu\nu}=e^{2\rho}\left(
  \begin{array}{cc}
    u^2-v^2 & v  \\
    v       & -1 \\
  \end{array}
\right)\>.\label{gParam}
\een
The variables $u$ and $v$ enter the action without time derivatives, and
act as Lagrange multipliers, enforcing the diffeomorphism constraints;
they are related to the shift and lapse
functions. The dynamical canonical fields are $\rho$, the
logarithm of the conformal factor, and $\eta$, the dilaton.
In first--order form, Eq.
(\ref{StrInsp}) becomes
\ben
S=\int dt\int d\sigma\> (\Pi_\rho\dot\rho+\Pi_\eta\dot\eta-\Ham)\>,
  \label{CanonAct}
\een
where an over--dot indicates derivative with respect to time $t$.
The Hamiltonian density $\Ham$ is a sum of constraints, expressed in
terms of canonical coordinates $\rho$, $\eta$, their
momenta $\Pi_\rho$, $\Pi_\eta$, and the Lagrange multipliers
$u$, $v$;
\ben
\Ham=u\E+v\P\>,\label{HamForm}
\een
where
\bea
\E&=&-2(\eta''-\rho'\eta')+\half\Pi_\rho\Pi_\eta+2\lambda e^{2\rho}
  \label{EConsA}\\
\P&=&-\Pi_\rho\rho'+\Pi_\rho'-\Pi_\eta\eta'\>.
  \label{PConsA}
\eea
We use a
prime to indicate differentiation with respect to the spatial
coordinate $\sigma$.

The action (\ref{CanonAct}) is equivalent to that
describing two free scalar and Hermitian
fields $r^a$, $\{a=0,1\}$, with indefinite metric
$\eta_{ab}={\rm diag}(1,-1)$ \cite{CJZ}.
We demonstrate this by making a canonical
redefinition from $\rho$, $\eta$, $\Pi_\rho$ and $\Pi_\eta$ to $r^a$ and
$\pi_a$ (at fixed time $t$, whose label is suppressed),
\footnote{The overlap matrix elements between the two sets of fields are
computed in \cite{Louis-Martinez}.}
\alpheqn\bea
\pi_0-\lambda r^{1\prime}&=&2\lambda e^\rho\sinh\Sigma
  \label{ScalDefA}\\
\pi_1+\lambda r^{0\prime}&=&-2e^\rho\cosh\Sigma
  \label{ScalDefB}\\
\lambda r^0&=&
 -\half e^{-\rho}(2\eta'\cosh\Sigma-\Pi_\rho\sinh\Sigma)
  \label{ScalDefC}\\
\lambda r^1&=&\half e^{-\rho}(\Pi_\rho\cosh\Sigma-2\eta'\sinh\Sigma)\>,
  \label{ScalDefD}
\eea\reseteqn
where
\ben
\Sigma(\sigma)=\half\int_{-\infty}^\sigma d\tilde\sigma\>
  \Pi_\eta(\tilde\sigma).\label{SigmaDef}
\een
In terms of the new fields $\pi_a$, $r^a$, Eq. (\ref{CanonAct}) becomes
\ben
  S=\int dt\int d\sigma\> (\pi_a\dot r^a-{\cal H})\>,\label{CanonAct2}
\een
with ${\cal H}$ as in (\ref{HamForm}), but now
\bea
\E&=&-\half\left(\frac{1}{\lambda}\pi^a\pi_a+
       \lambda r^{a\prime}r_a'\right)\nn\\
  &=&-\half\left(\frac{1}{\lambda}(\pi_0)^2+\lambda(r^{0\prime})^2\right)
     +\half\left(\frac{1}{\lambda}(\pi_1)^2+\lambda(r^{1\prime})^2\right)\nn\\
  &=&-\E^0+\E^1\>,\label{EConsB}\\
\P&=&-\pi_a r^{a\prime}=-\pi_0r^{0\prime}-\pi_1 r^{1\prime}\nn\\
  &=&\P^0+\P^1\>.\label{PConsB}
\eea
(Note the sign variation in $\E$; this leads to the variety of
vacua, whose properties we study in this paper.)

The theory defined by (\ref{CanonAct2})
has been quantized using the Dirac
procedure \cite{CJZ}, in a functional Schr\"odinger representation.
When we work with this formalism, by
letting the momenta $\pi_a$ act by functional
differentiation,
\ben
\pi_a\rightarrow\frac{1}{i}\frac{\delta\ \ }{\delta r^a}\>,\label{piDeriv}
\een
on wave functionals $\Psi(r^a)$ that depend on $r^a$, which act by
multiplication,
quantization consists of solving the constraint conditions
\ben
\E\Psi=\P\Psi=0\>.\label{ConsCond}
\een
There are two solutions $\Psi_\pm$,
\ben
\Psi_\pm=\exp\left[\pm i\frac{\lambda}{2}
  \int_{-\infty}^{\infty}d\sigma\> r^a\epsilon_{ab}r^{b\prime}\right]\>.
  \label{TwoSolns}
\een
Since a solution to the constraints exists explicitly, it
follows that the constraint algebra has no obstruction, and
satisfies the naive commutation relations, without center,
\alpheqn\ben
  i[\E(\sigma),\E(\tilde\sigma)]=i[\P(\sigma),\P(\tilde\sigma)]=
     \biggl(\P(\sigma)+\P(\tilde\sigma)\biggr)
         \delta'(\sigma-\tilde\sigma)\>,
   \label{EE_PP_comm}
\een
\ben
i[\E(\sigma),\P(\tilde\sigma)]=
     \biggl(\E(\sigma)+\E(\tilde\sigma)\biggr)
         \delta'(\sigma-\tilde\sigma)\>.
   \label{EP_comm}
\een\reseteqn
These commutators are obtained by applying the canonical commutation
relations
\ben
  i[\pi_a(\sigma),r^b(\tilde\sigma)]=\delta^b_a
     \delta(\sigma-\tilde\sigma)\label{pa_ra_comm}
\een
and ignoring issues of ordering.

A more careful analysis, which takes into account operator product
singularities, exposes the possibility of a center in (\ref{EP_comm})
\ben
i[\E(\sigma),\P(\tilde\sigma)]=
     \biggl(\E(\sigma)+\E(\tilde\sigma)\biggr)
         \delta'(\sigma-\tilde\sigma)-
         \frac{c}{12\pi}\delta'''(\sigma-\tilde\sigma)\>.
   \label{EP_comm_center}
\een
We may now understand why the above analysis, leading to the solution
(\ref{TwoSolns}), is not obstructed by a center. The indefinite signs of
the energy constraint (\ref{EConsB})
imply that $[\E,\P]=-[\E^0,\P^0]+[\E^1,\P^1]$, so
that if identical centers arise for both ``$0$'' and ``$1$'' fields,
they cancel.

We can write (\ref{CanonAct2}) in second
order form as the action of two free scalar fields.
The action $\barr S$,
\ben
\barr S=-\frac{\lambda}{2}\int dt\int d\sigma\>\sqrt{-g}g^{\mu\nu}
  \partial_\mu r^a\partial_\nu r^b \eta_{ab}\label{ScalAction}
\een
equals $S$ of (\ref{CanonAct2}) when the metric is parameterized
as in Eq. (\ref{gParam}), and $\barr S$
is put into canonical form.
The Hamiltonian
density is then identical to that of Eq. (\ref{HamForm}), with the
constraints given by Eqs. (\ref{EConsB},\ref{PConsB}).

When the gravitational theory is expressed as in (\ref{ScalAction}) it
resembles a bosonic string theory. Indeed if the model is defined on the
cylinder ${\bf R}^1\times S^1$, rather than on ${\bf R}^2$, we have a
closed bosonic string in $\{t,\sigma\}$ parameter space, propagating on
a flat two--dimensional target space $r^a$, with Minkowski metric tensor
$\eta_{ab}$. (It is interesting to note that the cosmological constant
$\lambda$ has become the string tension.)
As discussed in the introductory paragraphs,
the string theory cannot be quantized, due to the anomaly
in the constraint algebra (\ref{EP_comm_center}).
The anomaly in
string theory is insensitive to the signature of the target space,
and our indefinite metric would play no role. This is not in
contradiction with the fact that we found the solutions
(\ref{TwoSolns}), above. As shown in \cite{CJZ}, the anomaly is present
when the constraints are ordered with respect to the conventional string
theoretic vacuum, and absent when they are ordered with respect to the
functional Schr\"odinger vacuum.

In Section 2 we reformulate the gravity theory on ${\bf R}^1\times S^1$
and give a mode decomposition, so that the formalism coincides with that
of string theory. We review the normal ordering prescription (choice of
vacuum) used in string theory resulting in a center, and the one used in
the Schr\"odinger representation, which does not produce a center.

In Section 3 the ${\bf R}^1\times S^1$ theory is quantized with
Schr\"odinger--representation normal ordering; states analogous to
(\ref{TwoSolns}) are constructed, and further states are found, which
exist only on the cylindrical geometry. These give a representation for
the Virasoro algebra without center, which is absent with the chosen
ordering.

Section 4 is devoted to the string--type quantization. A center in the
Virasoro algebra exists; the operators are modified by the addition of
background charges, thereby increasing the center; ghost fields are
added with negative center so that the resulting total center
vanishes. BRST invariant states, which are deemed ``physical'', are
constructed.
The resulting spectrum is richer than what
is found in Section 3 \cite{Polyakov,BMP,LZ,Bilal};
however, the additional ``discrete'' states arise
at imaginary values of the zero--mode momentum, so the fields
in these theories transform
differently under Hermitian conjugation than those in Section 3. It is
the combination of the different Hermiticity properties of the fields
and the presence of background charges that is responsible for the
infinite tower of discrete states in the spectrum.
When we demand that the fields be Hermitian, we find
that there are no
discrete states: only the continuous portion of the BRST spectrum is
present. Consequently, we find fewer states with this approach than in
the functional Schr\"odinger quantization.

In Section 5, we consider the dilaton--gravity action (\ref{StrInsp})
coupled to a massless scalar field. This theory possesses a center for
either choice of ordering, arising from the matter degrees of
freedom \cite{CJZ}.
It was shown in \cite{CJZ} that in
the functional Schr\"odinger formulation the matter--coupled action is
equivalent to one studied in \cite{Kuch}, in which the theory was
quantized after modifying it to remove the center. In subSection 5.1 we
reproduce the argument of \cite{Kuch}, and present the spectrum. In
subSection 5.2 we quantize the theory
using the BRST approach, adopting a
string--like ordering for the fields. We introduce background
charges to increase the total, positive
center, which cancels against that of the ghost fields.
We then follow the development in
\cite{BMP,Bilal,Hirano} to compute the spectrum, again omitting states
with imaginary momentum. We find that
the resulting set of states
is larger than that presented in subSection 5.1, because of an additional,
unconstrained, zero--mode degree of freedom present in the BRST
spectrum.

In Section 6 we summarize and comment upon our results.

\mysection{Mode Analysis and Vacua}

With the action (\ref{ScalAction}) we work in conformal gauge, which we
fix by setting $u=1$, $v=0$ in the parameterized metric tensor
$g_{\mu\nu}$, given by (\ref{gParam}). The action then describes two
free scalar fields $r^a$, one entering with negative kinetic term, the
other with positive kinetic term (it is this alternation of sign that
leads to the variety of vacuum choices),
\ben
\barr S=\frac{\lambda}{2}\int dt\int d\sigma\>
  \left\{-\biggl((\dot r^0)^2-(r^{0\prime})^2\biggr)
  +\biggl((\dot r^1)^2-(r^{1\prime})^2\biggr)\right\}
     \>.\label{ScalFixed}
\een
The equations of motion
\ben
\frac{\delta\barr S}{\delta r^a}=0\Rightarrow
   \ddot r^a-r^{a\prime\prime}=0\label{raEOM}
\een
are solved by arbitrary functions of $t\pm\sigma$, and we apply spatial
periodic boundary conditions: the spatial interval is taken to be of
length $2\pi$, and henceforth we set $\lambda=1/4\pi$ for simplicity. We
expand the solution in terms of mode operators, consistent with the
periodicity requirement,
\ben
r^a(t,\sigma)=\xhat^a+2t\,\phat^a+i\sum_{n\neq 0}\frac{1}{n}\left[
  \alpha^a_n e^{-in(t-\sigma)}+\barr\alpha^a_n e^{-in(t+\sigma)}
 \right]\>.\label{raExp}
\een
The commutation relations (\ref{pa_ra_comm}) with
$\pi_a=-\frac{1}{4\pi}\dot r_a$, imply the algebra
\ben
  [\phat^a,\xhat^b]=i\eta^{ab}\label{pxCommRels}
\een
\ben
  [\alpha^a_m,\alpha^b_n]=
    [\barr\alpha^a_m,\barr\alpha^b_n]
    =-m\,\eta^{ab}\delta_{m+n,0}\>.
  \label{CommRels}
\een
The formal expressions for the constraints that would be
obtained by varying with respect to the multipliers $u$ and $v$, which
are now fixed, coincide with (\ref{EConsB}), (\ref{PConsB}). We express them in
terms of mode operators
\alpheqn\bea
L_m&\equiv&\half\int_0^{2\pi}d\sigma\>e^{-im\sigma}(\E+\P)\biggr |_{t=0}
  = -\eta_{ab}L^{ab}_m\label{VirOp}\\
\barr L_m&\equiv&\half\int_0^{2\pi}d\sigma\>e^{im\sigma}(\E-\P)\biggr |_{t=0}
  = -\eta_{ab}\barr L^{ab}_m\>,\label{BarVirOp}
\eea\reseteqn
\ben
L^{ab}_m=\half\sum_n :\alpha_{m+n}^a\alpha^b_{-n}:
   \quad ,\qquad
\barr L^{ab}_m=\half\sum_n :\barr\alpha_{m+n}^a\barr\alpha^b_{-n}:\>.
\label{VirMode}
\een
The colons denote an as yet unspecified ``normal'' ordering rule. Also
$\alpha^a_0=\barr\alpha^a_0\equiv\phat^a$. The structures associated
with the ``barred'' expressions are identical with the un--``barred''
ones, and the two commute with each other. But they do not act
independently of one another, since the zero mode oscillators
$\alpha^a_0$ and $\barr\alpha^a_0$ are identified.

The operators $L_m$ obey the Virasoro algebra
\ben
[L_m,L_n]=(m-n)L_{m+n}+\frac{c}{12}(m^3-m)\delta_{m+n,0}\>,\label{VirAlg}
\een
and similarly for $\barr L_m$. The
central element $c$ is determined by the normal--ordering
prescription. From this expression we see that
the constraint conditions
$L_m|\psi\rangle=\barr L_m|\psi\rangle=0$ are
consistent only if $c=0$.

We determine the center in the algebra (\ref{VirAlg}) by specifying a
vacuum state.
The convention adopted in string theory is that
the vacuum $\STvac{p^a}$
is annihilated by the positive--frequency mode operators,
\alpheqn\ben
\alpha^a_n\STvac{p^a}=\barr\alpha^a_n\STvac{p^a}=
0,\qquad n>0\>,\label{STvacDefA}\\
\een
and is an eigenstate of the zero--mode momentum operator $\phat^a$,
\ben
\phat^a\STvac{p^a}=p^a\STvac{p^a}\>.\label{STvacDefB}
\een\reseteqn
With this choice, the value of the center is independent of the
signature of the target--space
metric $\eta_{ab}$. The total center is therefore twice
that for a single scalar field,
$c=2$, and as a consequence the constraint conditions cannot be solved.
Furthermore,
the string vacuum $\STvac{p^a}$  also
gives rise to states of negative norm \cite{CJZ}.
This property makes it unsuitable for the functional Schr\"odinger
approach,
since in that formulation the norm of a state is given by a manifestly
positive functional integral.
Moreover, since $\alpha^0_{|n|}$ corresponds to
$\int_0^{2\pi}d\sigma\>e^{-i|n|\sigma}\left(\pi_0+
\frac{i|n|}{4\pi}r^0\right)$ (at $t=0$),
a state annihilated by $\alpha^0_{|n|}$ satisfies in the
Schr\"odinger representation the equation
\ben
i\int_0^{2\pi}d\sigma\>e^{-i|n|\sigma}\left(-\frac{\delta\ }{\delta r^0}
  +\frac{|n|}{4\pi}r^0\right)
   \Psi^{\rm vac}_{\rm ST}=0\quad ,\qquad n\not =0\>.\label{STFuncVac}
\een
This is solved by a quadratic exponential, which grows in function
space, and does not describe a localized, normalizable state.
($\Psi^{\rm vac}_{\rm ST}$ is the vacuum wave functional corresponding
to the abstract state $\STvac{p^a}$.)

In the functional Schr\"odinger formalism,
an alternative vacuum, $\FSvac{p^a}$, is adopted for the ``$0$''
variables with negatively--signed kinetic term.
This vacuum satisfies
\alpheqn\ben
\alpha^0_{-n}\FSvac{p^a}=\alpha^1_n\FSvac{p^a}=
\barr\alpha^0_{-n}\FSvac{p^a}=\barr\alpha^1_n\FSvac{p^a}=
   0,\qquad n>0\>,\label{FSvacDefA}
\een
\ben
\phat^a\FSvac{p^a}=p^a\FSvac{p^a}\>.\label{FSvacDefB}
\een\reseteqn
For the ``$0$'' excitations, the negative--frequency mode is taken to
annihilate the vacuum; since it contributes to the energy constraint
with a negative sign, in a sense it is equivalent to the
positive--frequency mode of the ``$1$'' excitations. Moreover, the
annihilation requirement, in contrast to (\ref{STFuncVac}), now demands
\ben
i\int_0^{2\pi}d\sigma\>e^{i|n|\sigma}\left(\frac{\delta\ }{\delta r^0}+
  \frac{|n|}{4\pi}r^0\right)
   \Psi^{\rm vac}_{\rm FS}=0\quad ,\qquad n\not =0\>,\label{FSFuncVac}
\een
which provides Gaussian, localized solutions.
($\Psi^{\rm vac}_{\rm FS}$ is the vacuum wave functional corresponding
to the abstract state $\FSvac{p^a}$.)
The solution for the
vacuum wave functional, satisfying (\ref{FSvacDefA}) and
the zero--mode condition (\ref{FSvacDefB}), which becomes
$\left(p_a+\int_0^{2\pi}d\sigma\>\pi_a\right)\Psi^{\rm vac}_{\rm FS}=0$, is
\ben
  \Psi^{\rm vac}_{\rm FS}=e^{-\frac{i}{2\pi}p_a\int_0^{2\pi}d\sigma\> r^a}
    \left[-\half\int_0^{2\pi}d\sigma\int_0^{2\pi}d\tilde\sigma\>
      (r^0\omega r^0+r^1\omega r^1)\right]\>,\label{FSFuncSoln}
\een
where
\ben
  \omega(\sigma,\tilde\sigma)=\frac{1}{8\pi^2}\sum_n|n|
     e^{in(\sigma-\tilde\sigma)}
   =-\frac{1}{16\pi^2}
     {\rm P}\,\frac{1}{\sin^2\half(\sigma-\tilde\sigma)}\>.
\een
With this choice, there are no negative--normed states; moreover, the
center is zero, because the one coming from the ``$0$'' oscillator
cancels against that of the ``$1$'' oscillator.
Evidently the spatial integral of the $\E$ operator is not positive
definite, because it vanishes on physical states by virtue of a
cancelation, but this is not a defect in our gravity theory, since
$\int d\sigma\E$ is not the energy of a ``physical'' state.

\mysection{Schr\"odinger--Dirac Quantization on the Cylinder}

For the cylindrical geometry,
we now construct states that are annihilated by the Virasoro generators,
which are ordered without a center, as in (\ref{FSvacDefA},b).
The constraints are most readily solved
by writing the Virasoro operators
as a product of factors. We shall display the calculations only for the
unbarred operators $L_m$; similar expressions can be constructed for the
barred ones.

We define light--cone combinations of the mode operators
\footnote{Our notation differs from that of \cite{CJZ}, where the
``$\pm$'' superscript refers to oscillators with positive and negative
kinetic term, while here we use it to denote light--cone combinations.}
\alpheqn\bea
\alpha^\pm_m&=&\frac{1}{\sqrt{2}}(\alpha^0_m\pm\alpha^1_m)
  \label{LCalpha}\\
\phat^\pm&=&\frac{1}{\sqrt{2}}(\phat^0\pm\phat^1)
  \label{LCphat}\\
\xhat^\pm&=&\frac{1}{\sqrt{2}}(\xhat^0\pm\xhat^1)\>,
  \label{LCxhat}
\eea\reseteqn
and find that they satisfy
\alpheqn\ben
[\alpha^\pm_m,\alpha^\pm_n]=0\>,\qquad
  [\alpha^+_m,\alpha^-_n]=-m\,\delta_{m,-n}\>,\label{LCcomms}
\een
\ben
[\phat^\pm,\xhat^\pm]=0\>,
   \qquad [\phat^\pm,\xhat^\mp]=i\>.\label{pxLCcomms}
\een\reseteqn
In terms of
these, the Virasoro operators appear in their factorized form,
\ben
L_m=-\sum_n :\alpha^+_{m+n}\alpha^-_{-n}:\>.\label{LCVirOp}
\een
{}From (\ref{LCcomms}) and (\ref{LCVirOp}),
we see that the normal--ordering symbol only affects
$L_0$, so the solutions $|\pm\rangle$ to the set of equations
\alpheqn\bea
\alpha^\pm_m|\pm\rangle&=&
   0\>,\qquad m\>\hbox{a nonzero integer}\label{pmKetA}\\
\phat^\pm|\pm\rangle&=&0\>,\label{pmKetB}
\eea\reseteqn
are annihilated by all of the Virasoro operators, with the possible
exception of $L_0$. That they are annihilated by $L_0$, as well, can be
verified directly or by applying the Virasoro algebra (\ref{VirAlg}),
with $c=0$.

The solutions to Eqs. (\ref{pmKetA},b) form two families ($\pm$),
labeled by the zero--mode momentum,
\ben
|\pm\rangle=\exp\left(\mp\sum_{n=1}^{\infty}\frac{1}{n}
  \alpha^1_{-n}\alpha^0_n\right)\FSvac{p^\mp}
  \equiv e^{\mp\Omega}\FSvac{p^\mp}\>,
 \label{PMexpr}
\een
where $\phat^\pm\FSvac{p^\pm}=p^\pm\FSvac{p^\pm}$,
and $\phat^\mp\FSvac{p^\pm}=0$.
This is annihilated by the $L_m$'s; since
a similar expression holds for the barred variables,
the solution to the full set of constraints, $L_m$ and
$\barr L_m$, is
\ben
e^{\mp\Omega}e^{\mp\barr\Omega}\FSvac{p^\mp}=
  \exp\biggl[
    \mp\sum_{n=1}^\infty\frac{1}{n}(\alpha^1_{-n}\alpha^0_n+
          \barr\alpha^1_{-n}\barr\alpha^0_n)
  \biggr]\FSvac{p^\mp}\>.\label{PMBARexpr}
\een
The two additional states $\Psi_\pm(r^a)=\langle r^a|\Psi_\pm\rangle$,
corresponding to (\ref{TwoSolns}), appear in terms
of oscillators as \cite{CJZ}
\ben
|\Psi_\pm\rangle=
  \exp\biggl[
    \mp\sum_{n=1}^\infty\frac{1}{n}(\alpha^1_{-n}\alpha^0_n-
          \barr\alpha^1_{-n}\barr\alpha^0_n)
  \biggr]\FSvac{0}\>.\label{TwoSolnsModes}
\een
There is a relative difference of a minus sign between the terms
depending on the barred and unbarred oscillators in (\ref{PMBARexpr})
and (\ref{TwoSolnsModes}). The negative sign preceding the barred
oscillators in (\ref{TwoSolnsModes}) can only appear when $p^a=0$.
The spectrum therefore contains a continuous component (\ref{PMBARexpr})
and a discrete component, consisting of the two states (\ref{TwoSolnsModes}).

The wave functionals that correspond to the solutions
(\ref{PMBARexpr}) can be obtained by writing the conditions
(\ref{pmKetA},b) in terms of the fields $r^a$,
\ben
  \langle r^a|e^{\mp\Omega}e^{\mp\barr\Omega}\FSvac{p^\mp}
   =e^{-\frac{ip^{\mp}}{2\pi}\int_0^{2\pi}
     d\sigma\>r^{\pm}}\>f^\mp(r^\pm),\label{PMBARfunc}
\een
where $f^\mp$ is independent of $r^\mp$ and is localized (by a
$\delta$--like functional) in $r^\pm$.

\mysection{BRST Quantization with Background Charges}

The constraint algebra (\ref{VirAlg}) with $c=2$ shows that a consistent
quantization condition for states built on the vacuum
$\STvac{p^a}$, defined by (\ref{STvacDefA},b), does not
exist because the center is nonvanishing.
Nevertheless, the corresponding
normal--ordering prescription has been used
to quantize theories that are closely related to (\ref{ScalFixed}) using
BRST quantization \cite{Polyakov,BMP,LZ,Bilal}. The center is removed by
adding background charges and ghost fields.
The resulting spectrum of ``physical'' states (the BRST cohomology)
is larger than the one that we found in the functional Schr\"odinger
formulation: there is a continuous component, corresponding to the
states (\ref{PMBARexpr});
there is also a discrete component, consisting of an
infinite tower of states at specific values of the zero--mode
momentum $p^a$. In a sense, these discrete states
are analogous to the two states
(\ref{TwoSolnsModes}); however, in BRST quantization, the zero--mode
momenta that label the discrete states
are imaginary when the background charges are real, so the fields
that enter the theory do not transform in a simple way under Hermitian
conjugation, in contrast to the Hermitian fields in
(\ref{ScalFixed}). Using the results of \cite{BMP}, and demanding that
the fields be Hermitian, we find that
only the continuous component of the
BRST spectrum is present, and there are no discrete states. More
precisely, there are states in the BRST cohomology that exist for
discrete imaginary values of the zero--mode momenta, but these states
are unphysical in our context
and we omit them from the spectrum.\footnote{
The states with imaginary momentum
appear to play a role \cite{Seib,Arisue} when the wave
functional of the universe is evolved using the Euclidean action
(as in the Hartle--Hawking construction \cite{HH}). Also,
for closely related reasons, such states arise when one attempts to cure
the non--normalizability of the string--theoretic vacuum by functionally
integrating over imaginary field configurations \cite{Arisue} (however,
see the discussion in \cite{Hirano}.)}
We now give the details of the above argument, 

We remove the obstruction in the Virasoro algebra
by adding background charges $Q^a$ to the
operators (\ref{VirMode}). We define new operators $L_m'$ and
$\barr L_m'$ (we record the formulae only for the unbarred
operators; similar expressions hold for the barred ones)
\alpheqn\bea
L_m'&=&-\half\eta_{ab}\sum_{n\not =-m,0}:\alpha^a_{m+n}\alpha^b_{-n}:
   -\phat_a\alpha^a_m-
   \,im\,Q_a\alpha^a_m\>,\quad m\not=0\>,\label{BCVirOp}\\
L_0'&=&-\half\eta_{ab}\sum_{n\not =0}:\alpha^a_{n}\alpha^b_{-n}:
   -\half \phat_a\phat^a-\half Q_aQ^a\>.\label{BCVirZero}
\eea\reseteqn
We note that this is not the form in which the constraints are
presented in \cite{BMP}. To make contact with that paper,
we define new momentum operators $P^a$
and background charges $\tilde Q^a$ by $P^a\equiv \phat^a-iQ^a$
and $\tilde Q^a\equiv -iQ^a$, in terms of which
(\ref{BCVirOp},b) become
\alpheqn\bea
L_m'&=&-\half\eta_{ab}\sum_{n\not =-m,0}:\alpha^a_{m+n}\alpha^b_{-n}:
   -P_a\alpha^a_m+
   \,(m+1)\tilde Q_a\alpha^a_m\>,\quad m\not=0\>,\label{BCVirOpAlt}\\
L_0'&=&-\half\eta_{ab}\sum_{n\not =0}:\alpha^a_{n}\alpha^b_{-n}:
   -\half P_aP^a+\,P_a\tilde Q^a\>.\label{BCVirZeroAlt}
\eea\reseteqn
The contribution of the ``$1$'' field to $L_m'$ is then the same
expression that is found in \cite{BMP} for the ``matter'' field. The
contribution from the ``$0$'' field, which enters with a negative sign,
can be identified with that of the ``Liouville'' field of \cite{BMP},
after making the change $\alpha^0_m\rightarrow i\alpha^0_m$:
despite the emphasis that we are placing on the role of
Hermiticity in determining the spectrum, this alteration
does not affect the conclusions
of this Section.

In the following we take $Q^a$ in (\ref{BCVirOp},b) to be
real. From (\ref{BCVirOp}) we see that $L_m'$ is Hermitian if the
oscillators satisfy $\alpha^{a {\rm\dag}}_m=\alpha^a_{-m}$, which also is the
condition appropriate for a Hermitian field. The spectrum that is found
in \cite{Polyakov,BMP,LZ,Bilal} contains states at imaginary eigenvalues of
the zero--mode momentum $\phat^a$. We shall therefore consider two cases:
$\phat^a$ anti--Hermitian, and $\phat^a$ Hermitian [the latter is the
appropriate one for the theory described by (\ref{ScalFixed})].
The Hermiticity
conditions that we apply to the Virasoro operators for
anti--Hermitian $\phat^a$ are \cite{Bilal} 
$L_m^{\rm \dag}(\phat^a)=L_{-m}(-\phat^a)$, while for Hermitian
zero--mode momentum they are
$L_m^{\rm \dag}(\phat^a)=L_{-m}(\phat^a)$.

In either case
we find, when we normal--order the modified Virasoro operators (\ref{BCVirOp})
using the string--type ordering,
that the center depends upon the
background charges $Q^a$,
\ben
c=2-12Q_aQ^a\>.\label{BCcenter}
\een
In BRST quantization, we add ghost fields $c_m$,
$\barr c_m$ and antighost fields
$b_m$, $\barr b_m$ to the theory, which contribute
$c_{\rm gh}=-26$ to the center. The BRST charge is then found to be
nilpotent only when the total center vanishes, so we fix $Q^a$ by setting
\ben
c_{\rm TOT}=-24-12Q_aQ^a=0\>.\label{BCGhcenter}
\een
The BRST cohomology for this theory has been computed \cite{BMP}.
It has a continuous component, consisting of two families of states,
labeled by the zero--mode momenta of the fields,
\ben
  |\psi\rangle_{\rm BRST}=\STvac{p^\pm}\otimes
      c_1\overline c_1|0\rangle_{\rm gh}\>.\label{BRSTcont}
\een
[The state $|0\rangle_{\rm gh}$ is the $SL(2,{\bf C})$ ghost vacuum.
The states $|\psi\rangle_{\rm BRST}$, which are annihilated by the
positive--frequency ghost and antighost oscillators,
are defined to have ghost number zero.]
These states correspond to the solutions (\ref{PMBARexpr}) that we found in
the previous Section, using the ordering appropriate to the
functional Schr\"odinger representation.

There is, in addition, a ``discrete'' component in the BRST spectrum
that has its counterpart in the two states
(\ref{TwoSolnsModes}). These additional
states appear at ghost number $0$, $\pm 1$
when the zero--mode momentum eigenvalues $p^\pm$ satisfy
\alpheqn\bea
p^+&=&irQ^+\label{pPlusCondI}\\
p^-&=&isQ^-\label{pMinusCondI}\>,
\eea\reseteqn
where $r$ and $s$ are integers, with $rs>0$. When the background
charge is nonzero, this gives an infinite tower of states labeled by
imaginary eigenvalues of the zero--mode momentum.

If the operator $\phat^a$ is anti--Hermitian, then its eigenvalues are
imaginary, and the discrete cohomology states are in the spectrum of the
theory. On the other hand, if $\phat^a$ is Hermitian, then we exclude
these states from the spectrum, which then contains only the continuous
family of states (\ref{BRSTcont}).
In the functional Schr\"odinger quantization performed in the previous
Section we found two discrete states at vanishing zero--mode
momentum. Since we did not need to introduce a background charge in that
approach, those states in a sense correspond to the case
$Q^\pm\rightarrow 0$ in (\ref{pPlusCondI},b). However,
in the BRST quantization, the case $r=s=0$ does not lead to a new state;
consequently, the BRST spectrum differs from the functional
Schr\"odinger spectrum both for Hermitian and anti--Hermitian $\phat^a$.

\mysection{Coupling to Scalar Matter}

In Sections 1-4 we studied the quantization of
the pure dilaton--gravity theory whose action is given
in Eq. (\ref{StrInsp}). We now introduce matter in the form of a
massless scalar field $f$ minimally coupled to the metric tensor
field,
\ben
S=\int dt\int d\sigma\>\sqrt{-g}\left[
  \frac{1}{4\pi G}(\eta R-2\lambda)+
  \half g^{\mu\nu}\partial_\mu f\partial_\nu f\right]\>.
  \label{GravMatt}
\een
Here $G$ is the gravitational coupling constant (``Newton's constant''),
and we continue to take
space--time to be a cylinder, so that $\sigma$ runs from $0$ to $2\pi$.
(The generalization to $N$ matter fields is straightforward but
unnecessary; we do not consider the large--$N$ limit.)
As stated in the Introduction, it is known that this theory cannot be
quantized without modification,
due to an obstruction \cite{CJZ}.
In subSection 5.1 we modify the action by a term that has
been considered previously \cite{CJZ,Kuch}, and
present a quantization of the modified theory
in the functional Schr\"odinger formalism, following
the treatment in \cite{Kuch}. In subSection 5.2 we follow the
development in \cite{Bilal,Hirano} to perform a BRST quantization of the
theory, using a string--like ordering for the gravitational fields, and
compare the different results of the two methods.

It is clear that (\ref{GravMatt}) may be reduced  to a sum of three free
scalar field actions, with indefinite kinetic terms. The reduction for
the gravity portion proceeds as previously described (with obvious
insertions of the factor $4\pi G$) while the matter already is in that
form.
In terms of the redefined gravitational
fields, (\ref{GravMatt}) in first order form becomes
\ben
S=\int dt\int d\sigma\>\left(
  \pi_a\dot r^a+\Pi\dot f
     -{\cal H}
  \right)\>,\label{GMCanonFF}
\een
where the matter variables
$\Pi$ and $f$ are canonically conjugate.
The Hamiltonian density is
${\cal H}=u\E+v\P$, with
\bea
\E&=&-\frac{1}{2}\left(
   G\pi_a\pi^a+\frac{1}{G}r_a'r^{a\prime}
  \right)+
  \half(\Pi^2+f^{\prime 2})
  \label{EConsD}\\
\P&=&-\pi_a r^{a\prime}-\Pi f'\>.
  \label{PConsD}
\eea
(As before, $\lambda$ has been set to $4\pi$.)
It is clear from these expressions that there is a center in the
constraint algebra. As discussed in Section 1, there is a contribution
to the center from each of the fields $r^0$, $r^1$, and $f$.
Those arising from $r^0$ and $r^1$ can
either add or cancel, depending upon the ordering that is
chosen. In either case, there is a nonvanishing center from the $f$
field, and
this remains true if we increase the matter content, since
each of the matter fields will contribute to the center with the same sign.

It is amusing to note that the above argument can be circumvented if we
adopt peculiar ordering conventions for each of the three fields
\cite{Niemi}, and in the Appendix we show that it is possible to
vary the center continuously by ordering in this fashion
with respect to a class of
generalized ``squeezed'' vacuum states.
However, our construction does not lead to a
well-defined quantum theory because of divergences that
continue to plague the formalism \cite{Jackiw}.

\subsection{Schr\"odinger quantization of the modified theory}

In this subSection we demonstrate that, by adding
terms dependent upon the gravitational fields
to the matter--coupled action (\ref{GMCanonFF}),
we can construct consistent quantum theories using the
functional Schr\"odinger formalism
\cite{CJZ,Kuch}. The
additional terms canonically
introduce a center in the
Poisson bracket algebra of the constraints that cancels the quantum
anomaly. Their
coefficients are proportional
to $\hbar$ (set
to unity in all of our expressions); in the classical limit, the
additional terms vanish, so the modified action corresponds in this
limit to the unmodified one (\ref{GMCanonFF}).

As a starting point, we use the first--order form
(\ref{GMCanonFF}) for the matter--gravity action, with the constraints
given in (\ref{EConsD}), (\ref{PConsD}).
We change variables in the
gravitational sector of the theory, from $\pi_a$, $r^a$ to new fields
$P_\pm$, $X^\pm$; in the new variables we can make rapid contact with the
results of \cite{Kuch}, which we shall use below
to quantize the modified
theory.

Up to this point it has not been necessary to specify the signs
of $G$ and $\lambda$; they occur in the combination $4\pi G/\lambda$,
and without loss of generality we had set $\lambda=4\pi$. (The results
for pure dilaton--gravity are insensitive to the sign of $\lambda$.)
However, below we shall argue that
in order to use the results of \cite{Kuch}, $4\pi G/\lambda$
should be negative, {\it i.e.} we take $G=-|G|$. The reasons for making this
choice become apparent later in the discussion.

The new fields are defined by the following transformation
\cite{CJZ,Kuch},
\alpheqn\bea
P_\pm&=&-\frac{\sqrt{|G|}}{2}(\pi_0+\pi_1)\mp
  \frac{1}{2\sqrt{|G|}}\left(r^{0\prime}- r^{1\prime}\right)\>,
  \label{PpmDef}\\
X^{\pm\prime}&=&\mp\frac{\sqrt{|G|}}{2}(\pi_0-\pi_1)-
  \frac{1}{2\sqrt{|G|}}\left(r^{0\prime}+ r^{1\prime}\right)\>.
  \label{XpmDef}
\eea\reseteqn
When we write (\ref{GMCanonFF}) in terms of
$P_\pm$ and $X^\pm$, it is convenient to express the action in terms of
constraints $C_\pm$, which are light--cone combinations of the constraints
$\E$, $\P$ defined in (\ref{EConsD}) and (\ref{PConsD});
\ben
C_\pm=-\half(\P\mp\E)=P_\pm X^{\pm\prime}\pm
  \frac{1}{4}(\Pi\pm f')^2\>.\label{CpmCons}
\een
With these definitions,
(\ref{GMCanonFF}) appears as
\ben
S=\int dt\int d\sigma\>\left[
 P_+\dot X^+ + P_-\dot X^- + \Pi\dot f
   -\mu_+ C_+ - \mu_- C_-
\right]\>,\label{KuActI}
\een
where
\ben
\mu_\pm=\pm u-v\>.\label{lamConsDef}
\een
The constraints $C_\pm$ satisfy a commutator algebra that is
determined by (\ref{CpmCons}),
(\ref{EE_PP_comm}), and (\ref{EP_comm_center}),
trivially modified to a quantization on a circle, rather than on an
infinite line.
\alpheqn\ben
i[C_\pm(\sigma),C_\pm(\tilde\sigma)]=
  -\biggl(C_\pm(\sigma)+C_\pm(\tilde\sigma)\pm\frac{1}{24\pi}\biggr)
  \delta'(\sigma-\tilde\sigma)\pm\frac{1}{24\pi}
  \delta'''(\sigma-\tilde\sigma)\label{Cpm_comm}\>,
\een
\ben
i[C_+(\sigma),C_-(\tilde\sigma)]=0\label{Cp_Cm_comm}\>.
\een\reseteqn
There is, as before, a center $c=1$,
coming solely from the matter fields,
and an additional contribution $\mp(1/24\pi)\delta'(\sigma-\tilde\sigma)$
beacuse we are quantizing on a circle, which was absent in Eq.
(\ref{EP_comm_center}) of Section 1, where we quantized on a line; unlike
the triple derivative term, the single derivative addition is trivial:
it can be removed by
adding $\mp 1/48\pi$ to the constraint operators.
The transformation (\ref{PpmDef},b)
to $X^\pm$, $P_\pm$, and use of the constraints
(\ref{CpmCons}), relies on the fact that the pure gravity portion of the
constraints is anomaly free: $P_\pm X^{\pm\prime}$ is like a field
theoretic momentum density, which has no anomalies in its algebra.

We now introduce the modified action $\tilde S$,
\ben
\tilde S=\int dt\int d\sigma\>\left[
 P_+\dot X^+ + P_-\dot X^- + \Pi\dot f
   -\mu_+ \tilde C_+ - \mu_- \tilde C_-
\right]\>,\label{KuActImod}
\een
where the new constraints $\tilde C_\pm$ are obtained from
$C_\pm$ by removing the trivial modification, and by adding a term proportional
to the center and dependent upon the gravitational ``coordinate''
fields $X^\pm$ \cite{CJZ,Kuch},\footnote{This is equivalent to adding a
background charge: see Ref.~\cite{CJZ}.}
\ben
\tilde C_\pm\equiv C_\pm\pm\frac{1}{48\pi}[\ln(\pm X^{\pm\prime})]''
     \mp\frac{1}{48\pi}\>.
  \label{CpmBarDef}
\een
The constraints $\tilde C_\pm$ then satisfy an algebra without center,
\ben
i[\tilde C_\pm(\sigma),\tilde C_\pm(\tilde\sigma)]=
 -\biggl(
  \tilde C_\pm(\sigma)+\tilde C_\pm(\tilde\sigma)
  \biggr)\delta'(\sigma-\tilde\sigma)\>,\label{Cpm_bar_comm}
\een
and, as a consequence, the Dirac quantization conditions
$\tilde C_\pm|\psi\rangle=0$ are consistent. However, we are
now faced with the problem of solving the constraint equations, which
include a complicated, non--polynomial term, dependent upon the
gravitational field variables.
Nevertheless, a solution to similar constraint equations was reported in
\cite{Kuch}; we adopt that argument for our purposes below.
\footnote{E.B. is grateful to
T. Strobl for enlightening
discussions about the work in \cite{Kuch}.}

The crucial observation of \cite{Kuch} is that the modified action
(\ref{KuActImod}) is canonically equivalent to one
in which the constraint
conditions have a simple form.
This canonical transformation
relies on an expansion of the matter fields
in terms of ``gravitationally dressed'' mode
operators $a^\pm_{n}$,
\ben
a^\pm_{n}\equiv\frac{1}{2\sqrt\pi}\int_0^{2\pi}d\sigma\>e^{inX^\pm}
  (\Pi\pm f')\>.\label{ainDef}
\een
(With this normalization, the dressed operators satisfy the usual commutator
algebra $[a^\pm_m,a^\pm_n]=m\delta_{m+n,0}$, $[a^+_m,a^-_n]=0$.)

Eq. (\ref{ainDef}) can be inverted,
and expressions for $\Pi$
and $f'$ obtained, when
the fields $X^\pm$ satisfy
\alpheqn\ben
X^\pm(2\pi)-X^\pm(0)=\pm 2\pi\>,\label{Xperiodic}
\een
and
\ben
X^{\pm\prime}\neq 0\qquad\mbox{everywhere.}\label{Xmonotonic}
\een\reseteqn
Together, these equations require that $X^+$ be monotonically increasing,
and that $X^-$ be monotonically decreasing. We show that this is
consistent with the combined transformations (\ref{ScalDefA}-d) and
(\ref{PpmDef},b). For $X^{+\prime}$ we find
\ben
X^{+\prime}=|G|^{-1/2}\exp\left(\rho-4\pi|G|\Sigma\right)>0\>,
  \label{XpMon}
\een
which is clearly consistent with (\ref{Xperiodic},b). The expression for
$X^{-\prime}$ does not have such a simple form; however, $P_-$ does:
\ben
P_-=-|G|^{-1/2}\exp\left(\rho+4\pi|G|\Sigma\right)\>,\label{PmMon}
\een
while from the expression (\ref{CpmCons}) for the unmodified constraints
we find that
$C_-=0$ implies
\ben
P_-X^{-\prime}=\frac{1}{4}(\Pi-f')^2\>,\label{PmXmMon}
\een
so that on the $C_-=0$ surface,
\ben
X^{-\prime}=-\frac{\sqrt{|G|}}{4} (\Pi-f')^2\,
  \exp\left(\rho+4\pi|G|\Sigma\right)<0\>.
  \label{XmMon}
\een
It is here that we see that we need to take $G$ negative.
Had we taken $G$ to be
positive, we would have found $P_->0$, which would then have implied
that $X^-$ is monotonically increasing, rather than decreasing, when
$C_-=0$. [The condition that the modified constraints vanish,
$\tilde C_-=0$, does not yield such a
straightforward result. However, we note that the added term
$-(1/48\pi)[\ln(\pm X^{-\prime})]''$ in (\ref{CpmBarDef}) is
singular at $X^{-\prime}=0$, so $X^{-\prime}$ should be strictly
positive or strictly negative, and it is reasonable that it should be
strictly negative, in agreement with the classical result (\ref{XmMon}).]

Inverting (\ref{ainDef}) we find
\bea
  \Pi&=&\frac{1}{2\sqrt{\pi}}
     \left(X^{+\prime}\sum_n e^{-inX^+}a^+_{n}-
     X^{-\prime}\sum_n e^{-inX^-}a^-_{n}\right)\label{Pi_exp}\\
  f'&=&\frac{1}{2\sqrt{\pi}}
     \left(X^{+\prime}\sum_n e^{-inX^+}a^+_{n}+
     X^{-\prime}\sum_n e^{-inX^-}a^-_{n}\right)\nn\\
  f&=&\frac{1}{2\sqrt{\pi}}
     \left(i\sum_{n\not = 0} \frac{1}{n}e^{-inX^+}a^+_{n}+
     i\sum_{n\not =0} \frac{1}{n}e^{-inX^-}a^-_{n}\>+
     X^+a^+_0 + X^-a^-_0+\xhat\right),\label{f_exp}
\eea
where $\xhat$ is independent of $\sigma$.
Because $f$ is periodic, $a^\pm_0$ are identified; we rename them $\phat$
since they are conjugate to $\xhat$,
\ben
\hat p\equiv a^+_{0}=a^-_{0}\>.\label{zero-modeDef}
\een

We now reorder the matter fields. This is convenient for further
development, and does not affect the center.
We order the fields
with respect to an eigenstate of $\phat$ that is annihilated by
the positive frequency gravitationally dressed mode operators,
\alpheqn\bea
\phat|p\rangle&=&p|p\rangle
  \label{KucVacA}\\
a^\pm_{n}|p\rangle &=& 0\>,\qquad n>0\>.\label{KucVacB}
\eea\reseteqn
This changes the constraint algebra to
\alpheqn\ben
i[\tilde C_\pm(\sigma),\tilde C_\pm(\tilde\sigma)]=
  -\biggl(\tilde C_\pm(\sigma)+F_\pm(\sigma)+\tilde C_\pm(\tilde\sigma)+
      F_\pm(\tilde\sigma)\biggr)
  \delta'(\sigma-\tilde\sigma)
  \label{CpmF_comm}
\een
\ben
F_\pm=\pm(1/48\pi)\left[4\sqrt{\pm X^{\pm\prime}}\left(
   \frac{1}{\sqrt{\pm X^{\pm\prime}}}
\right)''+(X^{\pm\prime})^2\right]\>,\label{Fpm_Def}
\een\reseteqn
and we must
introduce further terms in the modified constraints
(\ref{CpmBarDef}) to cancel
the new ordering--dependent contributions $F_\pm$.
The final constraints then read
\ben
\barr{C}_\pm=C_\pm\mp\frac{1}{48\pi}X^{\pm\prime}\left[
  X^{\pm\prime}+\left(\frac{1}{X^{\pm\prime}}\right)''
\right]\>,\label{NewKuCons}
\een
and they satisfy the algebra
(\ref{Cpm_bar_comm}), without center.

The dressed mode operators (\ref{ainDef}) all commute with
$\barr C_\pm$ in (\ref{NewKuCons}), and also with $X^\pm$. With this
information it is straightforward to verify that the following
transformation \cite{Kuch},
\alpheqn\bea
\barr X^{\pm\prime}\barr P_\pm&=&\barr{C}_\pm\label{KuchCOVA}\\
\barr X^\pm&=&X^\pm\label{KuchCOVB}
\eea
\bea
\frac{1}{2\sqrt{\pi}}
     \left[\barr X^{+\prime}\sum_{n\not =0} e^{-in\barr X^+}a^+_{n}-
     \barr X^{-\prime}\sum_{n\not =0} e^{-in\barr X^-}a^-_{n}
     +(\barr X^+ - \barr X^-)'\phat\right]&=&\Pi
   \label{KuchCOVC}\\
\frac{1}{2\sqrt{\pi}}
     \left[i\sum_{n\not =0} \frac{1}{n}e^{-in\barr X^+}a^+_{n}+
     i\sum_{n\not =0} \frac{1}{n}e^{-in\barr X^-}a^-_{n}\>+
     (\barr X^+ + \barr X^-)\phat+\xhat\right]
     &=& f\>,
   \label{KuchCOVD}
\eea\reseteqn
is canonical at the quantum level.

In terms of the new fields $\barr X^\pm$, $\barr P_\pm$,
the constraints take the simple form
$\barr{C}_\pm=\barr X^{\pm\prime}\barr P_\pm\>$, and
we can now quantize the theory in these variables
following the Dirac procedure.
Allowing the momenta
$\barr P_\pm$ to act by functional differentiation,
and $\barr X^\pm$ by multiplication, we see that
the condition that
$\barr{C}_\pm$ annihilate physical
states requires that the states are independent of $\barr X^\pm$. The
spectrum therefore consists of
$|p\rangle$, defined in (\ref{KucVacA},b), and states
constructed by acting on it with the
negative--frequency dressed operators $a^\pm_{-|n|}$,
subject to a further condition \cite{Kuch},
as we now explain.

The operators $a^+_{-|n|}$ and $a^-_{-|n|}$ cannot be applied
independently of one another, because of a constraint arising from the
periodic boundary conditions for the fields $r^a$: from
the transformation (\ref{PpmDef},b) that defines $X^\pm$, $P_\pm$, we
find that
$P_+-P_-=\sqrt{\Lambda}\left(r^{0\prime}-r^{1\prime}\right)$,
so the integral over the circle of $P_+-P_-$ vanishes,
\ben
 \int_0^{2\pi}d\sigma\>(P_+-P_-)=0\>.\label{PpmDifference}
\een
Evaluating
\ben
\int_0^{2\pi}\left(
  \frac{1}{X^{+\prime}}\barr{C}_+-
   \frac{1}{X^{-\prime}}\barr{C}_-
\right)=
\int_0^{2\pi}d\sigma\>(P_+-P_-)+
  \half\sum_{n\not=0}(:a_n^+a_{-n}^+:-:a_n^-a_{-n}^-:)\>,
\label{CpmDifference}
\een
we find that physical states $|\psi\rangle$ must satisfy
$\sum_{n\not=0}(:a_n^+a_{-n}^+:-:a_n^-a_{-n}^-:)|\psi\rangle=0$, which
relates the ``+'' and ``$-$'' oscillators appearing in the state.
This is the ``level--matching
condition'' familiar from string theory, which takes the above simple
form only after the reordering effected above---that is the reason for
reordering.

The solutions thus obtained are expressed in terms of the
gravitationally dressed mode operators. For some purposes this
representation is sufficient: in particular, in the next subSection, we
shall be able to compare directly the spectrum found above with that
obtained in BRST quantization. However, to make contact with the original
geometrical theory, it is desirable to represent the solutions
in terms of the variables $P_\pm$, $X^\pm$, $\Pi$, and $f$ that
enter the action (\ref{KuActI}). We now show that this cannot be done
without first solving the constraints (\ref{NewKuCons}) in terms of
those variables.

We find it convenient for this calculation to write the dressed mode
operators $a^\pm_{n}$ in terms of harmonic oscillator coordinates and
momenta $\Phi^\pm_{n}$ and $\Pi^\pm_{n}$ satisfying
$i[\Pi^\pm_{m},\Phi^\pm_{n}]=\delta_{mn}$,
\ben
a^\pm_{n}\equiv\frac{1}{\sqrt{2}}\biggl(
  \Pi^\pm_{|n|}-in\Phi^\pm_{|n|}
\biggr)\>.\label{HOCoords}
\een
We now compute the overlap matrix
$\langle f X^\pm|\Phi^\pm_{n}\barr X^\pm\rangle$ that relates the two
sets of canonical fields.
We obtain a set of functional differential equations
for the matrix elements from the equations defining the transformation,
(\ref{KuchCOVA}-d), by
promoting the fields to operators, and evaluating matrix elements of the
transformation equations. The solution is presented in terms
of a functional ${\cal M}$,
\ben
\langle f X^\pm|\Phi^\pm_{n}\barr X^\pm\rangle=
  \delta(\barr X^+-X^+)\delta(\barr X^--X^-)
  e^{
    i\int d\sigma\>[h^+h^{-\prime}-f'(h^+-h^-)]
   }\,{\cal M}
  \>,\label{KuchTrans}
\een
where
\ben
h^\pm=\sqrt{\frac{2}{\pi}}\left(
  \half\xhat+\sum_{n=1}^\infty\Phi^\pm_{n}\cos nX^\pm
\right)\>,\label{hpmDef}
\een
and ${\cal M}={\cal M}(X^+,X^-,f-h^+-h^-)$ satisfies
\ben
\left\{
  X^{\pm\prime}\frac{1}{i}\frac{\delta\ }{\delta X^\pm}\pm
   \frac{1}{4}\left(
      \frac{1}{i}\frac{\delta\ }{\delta f}\pm f'
   \right)^2
\mp\frac{1}{48\pi}X^{\pm\prime}\left[
  X^{\pm\prime}+\left(\frac{1}{X^{\pm\prime}}\right)''
\right]
\right\}{\cal M}(X^+,X^-,f)=0\>.\label{fCond}
\een
In order to determine the matrix elements we must solve the
equations for ${\cal M}$, but they are seen to be identical to
the condition that the
constraints $\barr{C}_\pm$ annihilate physical states.


\subsection{BRST quantization with background charges}

In this subSection we present a BRST quantization of the matter-coupled
theory.
We adopt a string-like ordering for the fields, so that the total center
from the gravitational and matter fields is $c=3$. We introduce
ghosts, which decrease the total center, and background charges, which
we adjust so that the center vanishes. Following the development in
\cite{Bilal,Hirano}, which relies upon the work of \cite{BMP},
we find a space isomorphic to the ghost-number zero states in the
cohomology of the BRST charge. However, our analysis
differs from that of \cite{Bilal,Hirano}; as with the
pure gravity theory, there are states in the cohomology
with imaginary eigenvalues of the zero-mode momenta.
They do not appear in our case, since we take the fields to be
Hermitian.
The spectrum then obtained has one more zero--mode
degree of freedom than that found in the functional Schr\"odinger
approach; otherwise, the spectra coincide.
(The fact that there are more
zero--mode degrees of freedom in the BRST states than the one associated
with the matter field was pointed out and discussed in \cite{CJZ}.)

Starting with the action (\ref{GMCanonFF}) we work in conformal
gauge, $u=1,~ v=0$.
To facilitate comparison with \cite{Bilal,Hirano}
we work with rescaled fields,
$r^a\to\sqrt{4\pi/|G|}r^a$, $f\to {\sqrt {4\pi}}f$.
(The sign of $G$ does not affect the calculations to be performed below;
however, to be consistent with the calculation in the previous
subSection we take $G$ to be negative, $G=-|G|$.)
As in Section 2 we expand the fields in modes
\alpheqn\bea
r^a(t,\sigma)&=&\xhat^a+2t\,\phat^a+i\sum_{n\neq 0}\frac{1}{n}\left[
\alpha^a_n e^{-in(t-\sigma)}+\barr\alpha^a_n e^{-in(t+\sigma)}
\right]\>,\label{raExpf}\\
f(t,\sigma)&=&\xhat^M+2t\,\phat^M+i\sum_{n\neq 0}\frac{1}{n}\left[
\alpha^M_n e^{-in(t-\sigma)}+\barr\alpha^M_n e^{-in(t+\sigma)}
\right]\>.\label{maExp}
\eea\reseteqn
We note that because of the choice of sign for $G$, now
$r^0$ has positive kinetic term, while that of $r^1$ is
negative. Consequently, the commutators of the mode operators differ
by an interchange $0\leftrightarrow 1$
from those of Section 2. They are now
\ben
  [\phat^a,\xhat^b]=-i\eta^{ab}\label{NewpxCommRels}
\een
\ben
  [\alpha^a_m,\alpha^b_n]=
    [\barr\alpha^a_m,\barr\alpha^b_n]
    =m\,\eta^{ab}\delta_{m+n,0}\>.
  \label{NewCommRels}
\een

We order with respect to an eigenstate $|p^a,p^M\rangle$ of the
zero-mode momentum operators that is annihilated by the positive
frequency mode operators,
\bea
\alpha^a_m|p^a,p^M\rangle&=&
\alpha^M_m|p^a,p^M\rangle=0,
\qquad m>0 \\
\phat^a|p^a,p^M\rangle&=&p^a|p^a,p^M\rangle,\\
\phat^M|p^a,p^M\rangle&=&p^M|p^a,p^M\rangle.
\eea
(We shall only record explicit expressions for the unbarred
degrees of freedom.) The Virasoro operators, modified by the
addition of background charges $Q^a$ to the gravitational fields, take
the following form,
\alpheqn\bea
L_m&=&\half\sum_{n\not =-m,0}(\eta_{ab}:\alpha^a_{m+n}\alpha^b_{-n}:
+:\alpha^M_{m+n}\alpha^M_{-n}:)\nonumber\\
&&\quad\qquad +\phat_a\alpha^a_m+\phat^M\alpha^M_m
+im\,Q_a\alpha^a_m\>,\qquad\qquad\quad m\not=0\>,\label{BCVirOpM}\\
L_0&=&\half\sum_{n\not =0}(\eta_{ab}:\alpha^a_{n}\alpha^b_{-n}:
+:\alpha^M_{n}\alpha^M_{-n}:)
+\half \phat_a\phat^a+\half(\phat^M)^2
-\half Q^aQ_a\>.\label{BCVirZeroM}
\eea\reseteqn
As in Section 4, we take the momenta to be Hermitian
and the background charges to be real, to be
consistent with the functional Schr\"odinger quantization of the
previous subSection. With the contribution from the background charges,
the center of the Virasoro algebra is given by
\ben
c=3+12Q_a Q^a.
\een
In BRST quantization, we add ghost fields $c_m$, ${\bar c}_m$ and
antighost fields $b_m$, ${\bar b}_m$ which contribute $c_{gh}=-26$ to
the center. Nilpotency of the BRST charge
is accomplished only when the total center
vanishes,
\ben
c_{\rm TOT}~~=~~c+c_{\rm gh}~~=~~-23+12Q_a Q^a~~=~~0.
\label{center}
\een
We use this condition to fix the value of $Q_aQ^a$.

Our task is now to find the ghost-number zero states in the cohomology
of the BRST charge $d_{\rm BRST}\equiv d+{\barr d}$,
which we identify as ``physical''. The unbarred fields
contribute to $d_{\rm BRST}$ the quantity $d$,
\ben
d=\sum^{\infty}_{n=-\infty} c_{-n} L_n -\half\sum^{\infty}_{m,n=-\infty}
:(m-n)c_{-m}c_{-n}b_{m+n}:.
\label{BRSTop}
\een
(The contribution $\barr d$ from the barred fields has an identical
form.) Since the barred and unbarred operators act independently (aside
from the level-matching condition, discussed at the end of this
subSection), it is sufficient to compute the cohomology of $d$, rather
than $d_{\rm BRST}$. 

In \cite{BMP}, a strategy for finding the $d$-cohomology was proposed.
It was shown that it is sufficient to compute the
cohomology of a much simpler operator. The appropriate operator emerges
after several steps. First, we define ${\hat d}$ as all of the terms in $d$
that do not contain the zero-mode ghost operators $c_0$ and $b_0$. We
find that $d$ can be written
\ben
d~~=~~c_0(L_0 +L_0^{gh})+b_0 M +{\hat d},
\label{DBRST}
\een
where $M$ contains only ghost fields, $L_0$ is defined in
(\ref{BCVirZeroM}), and $L_0^{gh}$ is the ghost Virasoro operator
$L_0^{gh}=\sum_{m}m :c_{-m}b_{m}:$. From (\ref{DBRST}) we see that $L_0
+L_0^{gh}= \{b_0,d\}$. A standard argument\footnote{
  Suppose that
  $|\psi\rangle$
  is an element of the $d$-cohomology
  and $(L_0+L_0^{gh})|\psi\rangle=h|\psi\rangle$, $h\not=0$.
  Then we can write $|\psi\rangle$ as
  $|\psi\rangle=d(b_0/h|\psi\rangle)$, which is a trivial
  ($d$-exact) state.} shows that
a nontrivial element of the cohomology must be annihilated by
\ben
L_0+L_0^{gh}\equiv p^+p^- +\half (p^M)^2-Q^+Q^- +{\hat L}-1=0,
\label{Emc}
\een
where ${\hat L}$ is the total ``level operator'' including gravity,
scalar matter and ghosts.
Next, we introduce the subspace ${\cal F}$ of states annihilated by
$L_0+L^{gh}_0$ and $b_0$,
\ben
{\cal F}=\{|\psi\rangle|(L_0+L^{gh}_0)|\psi\rangle=
0\ {\mbox {and}}\ b_0|\psi\rangle=0\}.
\label{Sub}
\een
{}From (\ref{DBRST}) we see that an element $|\psi\rangle$ of ${\cal F}$
annihilated by ${\hat d}$ is also annihilated by $d$; furthermore,
${\hat d}$ is nilpotent on ${\cal F}$, and we can consistently compute
its cohomology in that space.
It is possible to show \cite{BMP} that the
$d$-cohomology can be constructed from the ${\hat d}$-cohomology defined
on ${\cal F}$: to each $|\psi\rangle$ in the ${\hat d}$-cohomology
there correspond two possible elements, $|\psi\rangle$ and
$c_0|\psi\rangle$, in the $d$-cohomology.
However, we are interested in states with no ghost excitations,
and they are given by the ${\hat d}$-cohomology. 

We now introduce the operator whose cohomology is isomorphic to that of
${\hat d}$. Following \cite{BMP},
we assign ``degrees'' to the mode operators
\alpheqn\bea
{\mbox{deg}}(\alpha^+_n)&=&{\mbox {deg}}(c_n)=1,
\label{DegreeO}\\
{\mbox {deg}}(\alpha^-_n)&=&{\mbox {deg}}(b_n)=-1,\qquad (n\neq 0).
\label{DegreeT}
\eea\reseteqn
The degrees of all other operators are defined to be zero.
Then $\hat d$ is a sum of terms of degree $0$, $1$ and $2$,
${\hat d}={\hat d}_0 +{\hat d}_1 +{\hat d}_2$.
The contribution to ${\hat d}$ of zero degree is
\ben
{\hat d_0}=\sum_{n\neq 0}P^+(n)c_{-n}\alpha^-_n,
\een
where
\ben
P^+(n)=p^+ +iQ^+ n\>.
\label{Pmomentum}
\een
It was shown in \cite{BMP} that there is a one to one correspondence
between the ${\hat d}_0$ and ${\hat d}$ cohomologies (and also those of
${\hat d}_2$ and ${\hat d}$), so it is sufficient for our purpose to
compute the ${\hat d}_0$-cohomology, which is a relatively simple
problem.

We use a trick described in \cite{BMP} to find the states. We define the
operator
\ben
K\equiv \sum_{n\neq 0}{1\over P^+(n)}\alpha^+_{-n} b_n,
\label{inverse}
\een
which satisfies
\ben
\{ {\hat d}_0,K\}=\sum_{n\neq 0}(nc_{-n}b_n+\alpha^+_{-n}\alpha^-_n)
\equiv {\hat L}_{gg},
\label{Comm}
\een
where ${\hat L}_{gg}$ is the contribution to the level operator $\hat L$ from
the gravity and ghost fields.
(Note that $K$ is well defined for all real values of $p^+$ since
$P^+(n)$ never vanishes.)

Eq. (\ref{Comm}) implies that a state in the ${\hat d}_0$-cohomology is
annihilated by ${\hat L}_{gg}$, using the same argument that led us to
conclude that $L_0+L_0^{gh}$ [Eq. (\ref{Emc})] annihilates
nontrivial states in the
$d$-cohomology; consequently the nontrivial $\hat d_0$-cohomology states
have neither gravity nor ghost
excitations. Thus the ${\hat d}_0$-cohomology is the set of
all states constructed by acting an arbitrary number of times with the
negative-frequency matter oscillators $\alpha^M_{-|n|}$ on
the vacuum state $|p^{\pm},p^M\rangle$.
In addition, if the states are to lie in the subspace ${\cal F}$, as we
assumed, then they must also satisfy the
condition that they are annihilated by $L_0 +L_0^{gh}$, Eq. (\ref{Emc}).
(We note that from a theorem of \cite{BMP} we can, if we wish, construct
the explicit ${\hat d}$-cohomology states from those of
the ${\hat d}_0$ operator:  the procedure is explained in
\cite{BMP} and \cite{Hirano}.)

We have constructed the ghost-number zero states in the cohomology of the
unbarred operator $d$.
In order to obtain the
cohomology of the full BRST charge $d_{\rm BRST}$ we must 
consider the ${\barr d}$-cohomology as well. 
As previously mentioned, the ${\barr d}$-cohomology is just
a copy of that of the $d$ operator. In the $d_{\rm BRST}$-cohomology 
there is, however, an additional condition found by applying
$[(L_0+L^{gh}_0)-({\bar L}_0+{\bar L}^{gh}_0)]$ to physical states
$|\psi\rangle$. This is the level-matching condition,
\ben
\sum_n
(:\alpha^M_n\alpha^M_{-n}:-
:{\barr\alpha}^M_n{\barr\alpha}^M_{-n}:)|\psi\rangle=0.
\label{MachC}
\een
The physical states in the $d_{\rm BRST}$-cohomology are thus obtained by
applying the $\alpha^M_{-|n|}$ and
${\barr\alpha}^M_{-|n|}$ oscillators to $|p^\pm,p^M\rangle$, subject to the 
condition (\ref{MachC}).

At this point we note that had we allowed states with imaginary momenta
as in \cite{Bilal,Hirano}, we would have found a larger spectrum.
In that case, the operator $K$ is not always well defined since
there exist momenta $p^+$, and non-zero integers $n$, such that
$P^+(n)=0$. In the construction above, states with these momenta must be
treated as special cases; these are the discrete states.
However, we work with Hermitian fields, so we exclude the
discrete states from the spectrum.

We now compare the spectrum that we have obtained with that described in
subSection 5.1 using the functional
Schr\"odinger formalism. The latter states were obtained by acting
independently  with
an arbitrary number of the negative-frequency
dressed operators $a^+_{-|n|}$ and $a^-_{-|n|}$
of the scalar matter field on the
vacuum state, subject to the level matching condition. The spectrum is
therefore very similar to the one presented above. 
The BRST spectrum is also
constructed by applying two sets of negative-frequency creation operators,
$\alpha^M_{-|m|}$ and ${\barr\alpha}^M_{-|m|}$, again subject to 
the level-matching condition.
However, there is a difference between the two spectra
\cite{CJZ}.
The states obtained with the functional
Schr\"odinger formalism are labeled only by the zero-mode momentum of
the matter field, $p^M$.
In BRST quantization, we have three zero-modes $(p^+,p^-,p^M)$ and one
constraint, Eq. (\ref{Emc}), that restricts the momenta. 
The BRST states are thus labeled by two parameters,
which was also pointed out in \cite{CJZ}.
 

\mysection{Conclusion and Discussion}

In this paper we discussed two different approaches to quantizing the
string--inspired model of two dimensional gravity,
the functional Schr\"odinger and BRST methods. We treated both the pure
dilaton--gravity theory, and dilaton--gravity coupled to scalar matter.
Our main tool in this task was a sequence of field
redefinitions \cite{CJZ,Kuch}, which
we used to express the constraint conditions in alternative
forms, in which we recognized problems that have been discussed in the
literature \cite{CJZ,BMP,LZ,Bilal,Kuch}.
For both the matter--coupled and pure gravity theories we found
different spectra using the
two different quantization procedures.

In the case of pure gravity, there are two states in the functional
Schr\"odinger spectrum that have no counterpart in the BRST
approach (although in a sense they correspond to
the discrete states that appear at
imaginary momentum in the BRST cohomology); otherwise
the spectra coincide.

With the matter--coupled theory, there is a discrepancy in the
zero--mode degrees of freedom in the two spectra. The extra,
gravitational, zero--mode that appears in BRST quantization was
identified and commented upon in \cite{CJZ}. Its presence was considered
problematic, since the spectrum does not then resemble that of a
massless scalar field on a flat space--time, as one would expect from the
classical analysis. The functional Schr\"odinger spectrum, on the other
hand, obtained using the approach of \cite{Kuch},
does not share that difficulty, since there are no free
zero--modes in the gravitational degrees of freedom.
Consequently, for the matter--coupled
theory the functional Schr\"odinger
approach yields the most ``natural'' spectrum.

When we quantized the matter--coupled theory we argued that the
requirement that $X^+$ and $X^-$ be monotonically increasing and
decreasing, respectively [Eqs. (\ref{Xperiodic},b)],
could only be met when $4\pi G/\lambda$ is
negative.
As we emphasized in the body of the paper, the sign of $4\pi G/\lambda$
has no effect on any other calculation that we performed; however, it
does make a difference in the CGHS model \cite{CGHS}, from which our action
(\ref{StrInsp}) is
derived \cite{Kuns,CJZ},
since black hole solutions in the CGHS metric exist only for
positive cosmological constant.
On the other hand, we do not
expect that the restriction we encountered is generic,
for the following reason.
The constraint on the sign
arose from considering the field redefinitions
(\ref{ScalDefA}-d) and (\ref{PpmDef},b), and the form of those
transformations, in turn, is strongly constrained by the requirement
that the fields $r^a$ be periodic on the circle. The quantization that
we have performed on the circle should carry over (with obvious
modifications) to quantization on a finite interval, where more general
boundary conditions can be applied, and for which the restriction on the
sign of $4\pi G/\lambda$ need not hold. (A nice discussion of boundary
conditions is given in Ref.~\cite{Barvinsky}.)

Having obtained the physical states in the matter--coupled theory, a
natural next step is
to try to extract the space--time geometry
arising from a particular distribution of matter fields.
However, at present it is not possible to do this, because
we cannot explicitly
express the solutions in terms of the dilaton $\eta$ and the field
$\rho$ that we used to parameterize the metric tensor.\\[0.4in]

\noindent{\large\bf Acknowledgements}
\bigskip\medskip

\noindent
E.~B. thanks T.~Strobl for many useful discussions about the work
of K.~Kucha\v r.\\

\begin{appendix}
\bigskip\bigskip
\noindent{\large\bf Appendix}
\bigskip
\renewcommand{\theequation}{\mbox{A.\arabic{equation}}}
\setcounter{equation}{0}
\renewcommand{\alpheqn}%
{ \setcounter{saveeqn}{\value{equation}}%
  \stepcounter{saveeqn}\setcounter{equation}{0}%
  \renewcommand{\theequation}%
  {\mbox{A.\arabic{saveeqn}\alph{equation}}}}
\renewcommand{\reseteqn}%
{ \setcounter{equation}{\value{saveeqn}}%
  \renewcommand{\theequation}{\mbox{A.\arabic{equation}}}}

\noindent
We show that we can order the constraints
(\ref{EConsD}), (\ref{PConsD}) so
that their commutators satisfy an algebra without center.
We order the constraint operators with respect to a class
of vacuum states annihilated by linear combinations of
positive and negative frequency mode operators; that is, by linear
combinations of the usual ``annihilation'' and ``creation''
operators.\footnote{The same approach was used in \cite{Niemi} to
construct bosonic string theories in an arbitrary number of
target--space dimensions.} Such states are studied in quantum
optics, and are called ``squeezed states''.
We shall find, however, that states built upon these squeezed
vacua are not
invariant under finite action of the constraints, due to divergences
that arise when we order products of the constraint operators.

As in Section 2, we expand the fields $r^a$ and $f$ entering
(\ref{GravMatt}), and the constraints (\ref{EConsD}), (\ref{PConsD}) in
terms of mode operators. We work in conformal gauge, which we fix by
setting $u=1$ and $v=0$ in the parameterized metric tensor
(\ref{gParam}). We work with the rescaled fields of subSection 5.2,
$r^a\to\sqrt{4\pi/|G|}r^a$, $f\to {\sqrt {4\pi}}f$, which we expand in
terms of mode operators as in Eqs. (\ref{raExpf},b). (As discussed
previously, taking $G=-|G|$ to be negative implies that the ``0'' field
has positive kinetic term, while the ``1'' field has negative kinetic
term.) The Virasoro
operators have the form
\alpheqn\ben
L_m=\half\sum_{n\not =-m,0}(\eta_{ab}:\alpha^a_{m+n}\alpha^b_{-n}:
+:\alpha^M_{m+n}\alpha^M_{-n}:)
+\phat_a\alpha^a_m+\phat^M\alpha^M_m
\>,\ m\not=0\label{BCVirOpApp}
\een
\ben
L_0=\half\sum_{n\not =0}(\eta_{ab}:\alpha^a_{n}\alpha^b_{-n}:
+:\alpha^M_{n}\alpha^M_{-n}:)
+\half \phat_a\phat^a+\half(\phat^M)^2\>.\label{BCVirZeroApp}
\een\reseteqn
(A similar expression
holds for the barred operators $\barr L_m$: for the rest of this
Appendix, we present explicit expressions only for the unbarred operators.)
Both $L_m$ and $\barr L_m$ satisfy the
Virasoro algebra (\ref{VirAlg}) with a center that depends upon the
chosen operator ordering.

We now introduce the class of vacuum states that allows us to
vary continuously the value for the center. We define mode operators
$\talpha^a_m$, $\talpha^M_m$ by performing the
following transformation (a Bogoliubov transformation) for each of the
modes $m\not =0$,
\alpheqn\bea
\talpha^a_m&=&\alpha^a_m\cosh\theta_a+\alpha^a_{-m}\sinh\theta_a
  \label{tAlphADef}\\
\talpha^M_m&=&\alpha^M_m\cosh\theta_M+\alpha^M_{-m}\sinh\theta_M
  \label{tAlphMDef}
\eea\reseteqn
where $\{\theta_a,\theta_M\}$ are parameters that we take to be either real or
imaginary. We now define the state $\tvac$
\alpheqn\ben
\talpha^0_{|n|}\tvac=\talpha^M_{|n|}\tvac=0\>\label{tvac3DefA}
\een
\ben
\talpha^1_{-|n|}\tvac=0\>.
   \label{tvac3DefB}
\een\reseteqn
(Since our conclusions are unaffected by
the eigenvalues of $\phat^a$, $\phat^M$,
we take $\tvac$ to satisfy $\phat^a\tvac=\phat^M\tvac=0$ for simplicity.) 
The contribution to the center from each of the fields, when the Virasoro
operators are ordered with respect to $\tvac$, is conveniently computed
by substituting for $\alpha^a_m$ and $\alpha^M_m$ in terms of
$\talpha^a_m$ and $\talpha^M_m$ in (\ref{BCVirOpApp},b), and then evaluating
the commutator
\ben
[L_m-L_{-m},L_m+L_{-m}]=4mL_0+\frac{c}{6}(m^3-m)\>.\label{VirAlgShort}
\een
The center is found to depend upon the parameters $\theta_a$, $\theta_M$,
\ben
c = \cosh 2\theta_0-
   \cosh 2\theta_1+\cosh 2\theta_M\>.\label{BogCent1}
\een
(Note that this gives $c=1$ when
all of the $\theta$'s vanish, as expected.)
It is clear that there are many solutions $\{\theta_a,\theta_M\}$ for which
$c=0$.

With the center set to zero it is possible in principle to solve
the constraint equations $L_m|\psi\rangle=0$
for a state $|\psi\rangle$ built upon the vacuum $\tvac$
defined by Eqs. (\ref{tvac3DefA},b).
There is a difficulty, however,
with the interpretation of the solution obtained in this way.
In \cite{Jackiw}, a calculation related to ours suggests
that the Virasoro operators
ordered with respect to vacua defined by (\ref{tvac3DefA},b)
cannot be exponentiated to yield finite transformations,
because of infinities that appear when products of the
constraint operators are ordered with these states.\footnote{
Moreover, in \cite{Niemi} it was argued that, from the point of view of
string theory, the excitations
of the string cannot be interpreted as particle states.}
Below, we show how these infinities arise in our calculation.
We shall find that while we can cancel them at quadratic order,
new divergences arise at higher orders.

When we evaluate the vacuum expectation value of the square of the
Virasoro operator $L_m$ defined in (\ref{BCVirOpApp}), we find
\ben
\widetilde{\langle 0|}L_m^2\tvac=C\left[
  \frac{1}{12}|m|(m^2-1)+\sum_{n=1}^\infty (|m|+n)n
\right]\>.\label{InfSqrd2}
\een
where $C=\frac{1}{4}(\sinh^2 2\theta_0+\sinh^2 2\theta_1+\sinh^2 2\theta_M)$.
The second of the bracketed terms in (\ref{InfSqrd2}) is divergent,
and for real
values of the parameters each of the fields contributes with the same
sign. However, for imaginary
$\theta$, $\sinh^22\theta$ is negative, and we can arrange a
cancelation in the prefactor $C$ by taking some of the parameters to be
imaginary, and others to be real.
(We also note that the contribution to the
center, $\cosh 2\theta$, is real for imaginary $\theta$.)
We demonstrate that
it is possible to solve simultaneously
the conditions of vanishing center and vanishing vacuum expectation
value of $L_m^2$ by recording a particular solution,
\alpheqn\ben
\theta_0=\theta_M=\frac{i}{2}\beta\>,\quad
  \cos\beta=\frac{1}{\sqrt{2}}\label{ThetaSolnA}
\een
\ben
\cosh 2\theta_1=\sqrt{2}\>.\label{ThetaSolnB}
\een\reseteqn
However, the cancelation does not persist to higher orders: evaluating
$\widetilde{\langle 0|}L_m^4\tvac$ we find
\ben
\widetilde{\langle 0|}L_m^4\tvac=
3(\widetilde{\langle 0|}L_m^2\tvac)^2+
 C_1\sum_{n=1}^\infty (|m|+n)^2(2|m|+n)n+
 C_2\sum_{n=1}^\infty (|m|+n)^2n^2+{\rm finite}
   \>,\label{InfQuart}
\een
where the coefficients $C_1$ and $C_2$ are
given by $C_i\equiv c_i(\theta_0)+c_i(\theta_1)+c_i(\theta_M)$, with
\alpheqn\bea
c_1(\theta)&=&\left[
  12(\cosh^4\theta+\sinh^4\theta)\sinh^2 2\theta
     +\frac{9}{8}\sinh^4 2\theta\right]\label{C1Def}\\
c_2(\theta)&=&3\sinh^4 2\theta\>.\label{C2Def}
\eea\reseteqn
We see that $c_2$ is positive even for imaginary $\theta$; also,
$c_1$ will not vanish in general, when the $\theta$'s are chosen
such that (\ref{InfSqrd2}) vanishes. Moreover, it is reasonable to
expect that we shall encounter further divergences at higher orders.

\end{appendix}

\pagebreak

\end{document}